\documentstyle[aps,prb,epsf,multicol]{revtex}

\newcommand {\fd} {\rightarrow}

\newcommand {\bay} {\begin{array}}
\newcommand {\eay} {\end {array}}
\newcommand {\be}{\begin{equation}\noindent}
\newcommand {\ee}{\end{equation}\noindent}
\newcommand {\ba}{\begin{eqnarray}}
\newcommand {\ea}{\end{eqnarray}}
\newcommand {\bd}{\begin{displaymath}}
\newcommand {\ed}{\end{displaymath}}
\newcommand {\bi}{\begin{itemize}}
\newcommand {\ei}{\end{itemize}}

\newcommand {\formu}[1] {(\ref{#1})}
\newcommand {\tavola}[1] {Tab.~\ref{#1}}
\newcommand {\sezione}[1] {Sec.~\ref{#1}}

\newcommand {\fig}[1] {Fig.~\ref{#1}}

\newcommand {\bn} {\begin{enumerate}}
\newcommand {\en} {\end{enumerate}}
\newcommand {\bc} {\begin{center}}
\newcommand {\ec} {\end{center}}

\begin{document}
\draft
\title{Against Temperature Chaos in Naive TAP Equations}
\author{R. Mulet$^1$, A. Pagnani$^2$, G. Parisi $^3$}
\address{$^1$ Abdus Salam ICTP, Strada Costiera 11, P.O. Box 586,I-34014 Trieste, Italy.\\
$^2$ Dipartimento di Fisica and INFM,
Universit\`a ``Tor Vergata'',
V. della Ricerca Scientifica 1, I-00133 Roma.\\
$^3$ Universit\`a  ``La Sapienza'', INFN and INFM Unit\`a di Roma,
P. Aldo Moro 2, I-00185 Roma.\\
}
\date{\today}
\maketitle

\begin{abstract}
We study the temperature structure of the naive TAP equations by mean
of a recursive algorithm. The problem of the chaos in temperature is
addressed using the notion of the temperature evolution of equilibrium
states. The lowest free energy states show relevant correlations with
the ground state, and a careful finite size analysis indicates that
these correlations are not finite size effects, ruling out the
possibility of chaos in temperature even in the thermodynamic
limit. The correlations of the equilibrium states with respect to the
ground state are investigated. The performance of a new heuristic
algorithm for the search of ground states is also discussed.
\end{abstract}

\section {Introduction}

The mean-field theory of spin glasses (SG) based on the
Sherrington-Kirkpatrik (SK) model has revealed very interesting
properties of its low-temperature phase. Among them there are a rugged
free-energy landscape with many metastable states, universality of the
probability distribution function of the overlap between states, and
their ultrametric organization \cite{BOOKS}. These properties
are common to various randomly frustrated systems, but the full
analytic control is formulated mostly for the SK model.

The first attempt toward a Curie-Weiss theory of the SG phase has been
done within the TAP theory~\cite{TAP}, in which a set of non-linear
equations for the mean site magnetizations $\{m_i\}$ has been
introduced much in the same spirit of the mean-field equations for
ordinary ferromagnets. Such a set of equations has been analytically
investigated in~\cite{BRAY_MOORE}, where it has been found that the
number of minima increases exponentially with the system size
(precisely as $\exp(\alpha(T)N)$, being $\alpha(T)$ some
temperature dependent $O(1)$ constant and $N$ the number of spins),
and that there exists a temperature dependent free energy threshold
above which the solutions of the TAP equation are {\em uncorrelated}
(i.e. their mutual overlap is zero in the limit of
$N\fd\infty$). However a number of questions about the detailed
structure of the metastable states of the SK model still remains
unanswered.

A direct numerical solution of the TAP equations is difficult because
of the so called Onsager reaction term (we will discuss about it in
the following), which introduces a portion of the configuration space
in which the equations themselves lose their validity
\cite{BRAY_MOORE,NEMO_TAKA_85}. Here we will study a {\em simplified}
version of TAP equations, obtained from the original ones dropping the
numerically {\em dangerous} Onsager reaction term. This new set of
equations is known as the {\em naive} mean-field (NMF) equations, and
they become the exact mean-field equations for a generalized SK model
\cite{BRAY_SOMPO_YU}. This model turns out to have strikingly similar
SG properties to the original SK model \cite{NEMO_TAKA_90}, capturing
all the complexities of the SG phase, but with mean-field equations
open to easier numerical integration. This model, as well as the
original SK model, displays Replica Symmetry Breaking.

The main focus of this work has been the analysis of the organization
of the equilibrium states at different temperatures. The motivation
that led us to this problem is that while, at least at the mean-field
level, it is well known that below $T_C$ states at the same
temperature in the SG phase are non-trivially correlated, very little
is known about the correlations between states at different
temperatures, despite intense theoretical and numerical efforts on
this subject in the last fifteen years.

The hypothesis of the chaos in temperature can be very simply stated
in terms of absence of correlations between states at different
temperatures. It was originally introduced as a constitutive
ingredient of the phenomenological droplet theory
\cite{FISHER_HUSE_86} in order to take into account the absence of
strong cooling rate dependence experimentally not observed in real
spin glasses \cite{BRAY_MOORE_87,FISHER_HUSE_88}: the approach to
equilibrium of a given observable after cooling from the high
temperature to a working temperature $T_W<T_C$ does not depend on the
thermal history of the experimental sample, but only on the time spent
at the last temperature $T_W$ \cite{EXPERIMENTS}. The puzzle becomes
more intricated once we turn our attention to the memory effects
observed in temperature cycling experiments \cite{EXPERIMENTS}: the
state reached by a system at a given temperature below $T_C$ is
recovered after a negative temperature cycle. The memory effects are
manifestly contradicting the chaotic scenario suggested by the cooling
rate insensitivity, giving rise at the same time to a number of
theoretical explanations mostly focused on real space point of view
\cite{REAL_SPACE1,REAL_SPACE2}. Let us stress a possible source of
misunderstanding: in \cite{BRAY_MOORE_87} it was related temperature
chaos with bond-chaos, i.e. perturbation on the systems induced by
infinitesimal changes in the quenched disorder. While the latter it
has been clearly shown to be present \cite{RIEGER}, we believe to be
able to demonstrate in what follows that the two forms of chaos are
different and that, at least at the mean field level, no temperature
chaos is present.

On the pure theoretical side, even at the mean-field level, the
situation is still far from being satisfactory. Chaos in temperature
was first advocated in an unpublished work of Sompolinsky, then
reconsidered negatively in \cite{KONDOR_1} at zero-loop order
(i.e. for the infinite range limit of the theory), then again
supported in \cite{KONDOR_2} at one-loop order (i.e. for the short
range case). Lastly we point out the work of Franz et
al. \cite{FRANZ_NEY-NIFLE} where, by means of the coupled real
replicas method \cite{FRA_PA_VI}, they supported the presence of
chaos. This approach was recently reconsidered in \cite{TESI_TOMMASO},
where it was demonstrated that taking into account higher order
perturbation terms, there is no chaos in temperature.

On the numerical side, clues of chaos in temperature for short range
spin glass model were discussed in \cite{FELIX,NEY,NEY_YOUNG}, while
our results are in substantial agreement with the simulations
presented in \cite{KISKER} where it was observed that the overlap
correlation length in 3D Edwards-Anderson model (EA) is a
monotonically increasing length scale with respect of the aging time,
and with the more recent simulations of the SK model and the 3D
Edwards-Anderson model presented in \cite{BILL_MARI}.

In this work we carried out a careful numerical analysis of the
solutions of the naive TAP equations. We solved these equations by
using a recursive algorithm in the spirit of
\cite{NEMO_TAKA_90,LA_MA_PA}.  We managed to give a coherent
description of the temperature structure of the solutions, that we
have classified into two families: solutions appearing just below the
critical temperature which display a bifurcation scenario as the
temperature is decreased, and a huge number of solutions appearing
well below the critical temperature. We have addressed our
investigation on the changes in temperature of the free energy
landscape by operatively defining the temperature evolution of a
generic equilibrium state. This allowed us for the analysis of the
correlations between the ground state and its temperature evolved
state: the obtained results suggest a temperature smoothly varying
free energy landscape, and consequently a non chaotic
scenario. Exploiting this temperature evolution technique we have also
been able to characterize the nature of the first solutions just below
$T_C$: it turns out that these solution are highly correlated with the
ground state, suggesting a scenario in which the first minima
appearing just below the paramagnetic phase are also the deepest ones
throughout the whole SG phase. Let us stress that our findings are
strengthened by a careful finite size scaling analysis that rules out
the possibility that our results are finite size effects. The
interpretation of the temperature structure of the equilibrium phase
space has suggested to us the implementation of a heuristic algorithm
for the search of the ground states that gives an approximate value
for the ground state energy density which is always less than $1\%$
higher than the true ground state.

The paper is organized as follows. In \sezione{sec_model} we will
introduce both the model and the numerical method. In
\sezione{sec_tree} we will discuss the nature of the tree of solutions
of the naive TAP equations in terms of their free energy landscape. In
\sezione{sec_org} we will discuss the temperature organization of the
equilibrium states. We will show in terms of free energy landscape
that the minimum originating from the ground state shifts smoothly
with respect of temperature changes with no sign of temperature
chaos. In \sezione{sec_algo} we will discuss the performance of a new
heuristic algorithm for the search of the ground states. In
\sezione{sec_concl} we will briefly summarize our findings and comment
on further developments.

\section{Model and numerical methods}
\label{sec_model}
The main difference between the replica method and the TAP approach is
that while in the former we can only calculate observables averaged
over the disorder probability distribution function, using the latter
we can gain information about single sample quantities, eventually
averaging the results over the disorder at a later stage of the
calculation. A rigorous derivation of the self-consistency equations
for the magnetization of the SK model gives \cite{BOOKS,TAP}:
\begin{equation}
\label{eq_tap}
  m_i = \tanh\left[\beta \left(\sum_{j=1}^N J_{ij}m_j + h_i - \beta
  \sum_{j=1}^N J^2_{ij}(1-m_j^2)m_i\right)\right]\,.
\end{equation}
This formula is a system of $N$ coupled non-linear equations for the
local magnetizations $m_i \equiv \langle s_i\rangle$. The physical
interpretation of the terms involved in the TAP equations is rather
simple: the first two terms inside the hyperbolic tangent in the r.h.s
of \formu{eq_tap} are the usual mean-field terms of ordinary
ferromagnets, while the last one is the Onsager reaction term, which
is a measure of the contribution to the internal field acting on the
site $i$ coming from the magnetization $m_i$ itself. Such a term is
relevant in the case of the SK model because the generic coupling
$J_{ij}\propto N^{-1/2}$, while for a ferromagnetic theory the
coupling is $O(1/N)$ (the free energy must be proportional to $N$), so
that, at the mean-field level, this term is never taken into
account. In the paramagnetic phase and for small values of $\tau =
T_C-T$ the local magnetizations are expected to behave as $m_i =
\tau^{1/2} + O(\tau)$, so that the Onsager term drops to zero as
$\tau^{3/2}$ apart from a constant term whose main effect is to shift
the critical temperature. In the low temperature phase ($\beta \fd
\infty$) a detailed replica calculation shows that $q_{EA}\equiv
1/N\sum m_i^2= 1 - O(\beta^{-2})$ (at least for the lowest energy
states), so that $\beta( 1 - q_{EA}) = O(\beta^{-1})$ and again the
Onsager term drops. The analytical calculation presented in
\cite{NEMO_TAKA_90} and numerically confirmed in
\cite{NISHI_NEMO_TAKA}, yields a value of $T_C=2$ for the critical
temperature, where it has to be noticed that the critical temperature
for the SK model (i.e. for the complete TAP equation \formu{eq_tap})
is $T_C=1$.

In the following we will set to zero the Onsager term and the extern
magnetic field $h_i$, and we will use random symmetric Gaussian
couplings of zero mean and variance $1/N$. Under these assumptions we
can rewrite \formu{eq_tap} as follows:
\begin{equation}
\label{eq_ntap}
   m_i = \tanh\left(\beta \sum_{j=1}^N J_{ij}m_j \right)\,.
\end{equation}
As we have mentioned above in the low temperature phase
\formu{eq_ntap} has a lot of solutions. We shall label them with
$m_i^\alpha$, and we shall use in general the superscript $\alpha$ to
denote quantities such as the free energy $F^{\alpha}$ related to the
corresponding solution:
\begin{equation}
\label{eq_free_energy}
F^{\alpha} = E^{\alpha} - \frac {S^{\alpha}}{\beta}\,,
\end{equation}
where the internal energy $E^\alpha$ and the entropy $S^\alpha$ are
given by:
\begin{eqnarray}
E^{\alpha} &=& - \sum_{1\leq i < j\leq N}J_{ij}m_i^\alpha m_j^\alpha \\
S^{\alpha} &=& - \frac{1}{2} \sum_{i=1}^N \left[ (1+m_i^\alpha)\ln
\left(\frac{1+m_i^\alpha}2\right) + (1-m_i^\alpha)\ln
\left(\frac{1-m_i^\alpha}2\right)\right]\,.
\end{eqnarray}
In the following we will mainly deal with densities of thermodynamic
potentials. We shall adopt for these quantities the standard lowercase
notation $f^\alpha,e^\alpha, s^\alpha$.

Let us stress that, at least for the naive TAP equations, it is easy
to demonstrate that the critical temperature $T_C$ is self-averaging
observing that in the paramagnetic phase (i.e. high temperature) the
equation \formu{eq_ntap} can be rewritten as $m_i = \beta \sum_k
J_{ij}m_j$. The transition temperature $T_C$ in this contest is the
temperature where the paramagnetic solution $\{m_i=0\}$ becomes
unstable. From random matrix theory \cite{MEHTA} it is known that this
temperature corresponds with the higher positive eigenvalue (which is
$2$) and moreover such an eigenvalue turns out to be self-averaging.

Denoting with $N_S(T)$ the number of solutions of \formu{eq_ntap} at
temperature $T$, the link between the TAP formalism and the single
sample mean-field partition function is given by
\begin{equation}
\label{eq_z}
   Z_{J} = \sum_{\alpha=1}^{N_{S}(T)} e^{-\beta F^\alpha}\,,
\end{equation}
which encodes the intuitive idea of Boltzmann weighted sum over the
solutions \cite{DE_DOM_YOUNG}. The internal density of energy can
be expressed, according to \formu{eq_z}, as:
\begin{equation}
e = -\frac1V \frac{\partial\ln(Z_J)}{\partial \beta} = 
\frac1V \sum_{\alpha=1}^{N_{S}(T)} \frac{e^{-\beta F^\alpha}}{Z_{J}}
\frac{\partial }{\partial \beta} (\beta  F^\alpha) =
\sum_{\alpha=1}^{N_{S}(T)} e^\alpha w^\alpha\,,
\end{equation}
where we defined the weights $w^\alpha \equiv e^{-\beta
F^\alpha}/Z_{J}$, and we exploited the stationarity conditions
$\partial_{m_i} F = 0$, which are the mean-field equations
\formu{eq_ntap}.

In order to solve \formu{eq_ntap} we used a recursion algorithm
in the spirit of \cite{NISHI_NEMO_TAKA,FIRST_RECUR}. We start by
generating a random initial configuration $m_i^{(0)}$ taken from a
flat probability distribution function with support in $[-1,1]$, and
then we start the recursion:
\begin{equation}
\label{eq_itera}
 m_i^{(t+1)} = \tanh\left(\beta \sum_{j=1}^N J_{ij}m_j^{(t)}\right)\,.
\end{equation}
The algorithm stops when $\sum |m_i^{(t+1)}- m_i^{(t)}|< 10^{-6}$. The
algorithm was implemented in double precision, in order to avoid
rounding problems. We also checked that the state obtained by the
converging recursion is effectively a solution of \formu{eq_ntap} by
direct inspection. In the following we will assume that two solution
$\{m_i^\alpha\}, \{m_i^\beta\}$ of \formu{eq_ntap} have to be
considered as different whenever $\sum |m_i^\alpha - m_i^\beta | >
10^{-2}$. Given this requirement we have found no difficulty in
distinguishing solutions. We have also checked that the
results do not change when more stringent bounds are adopted.  Note
that the maximum volume reached in our simulations was $N=200$.

The exhaustive enumeration of the solutions at a given temperature was
achieved following this simple procedure: each solution obtained by
the iteration is accumulated into a stack if it is different from all
the other solutions already accumulated. Each solution is also
labelled with the number of times that it has been found by the
recursion. The search stops after all solutions in the stack have been
found at least $10$ times. We also tried to increase this number but
we did not find any increase in the total number of the solutions, at
least within the volume accessible for exhaustive enumerations, which
for us is $N\leq 28$ for the whole SG phase, and $N\leq 200$ down to
$T=T_C/2=1$. We insist here on the strict requirement of exhaustive
enumeration, by checking that each solution has the magnetic reversal
counterpart. It is in fact obvious that if $\{\hat m_i$\} is a
solution of \formu{eq_ntap}, also $\{-\hat m_i\}$ will be a solution
with the same free energy as a consequence of the overall $Z_2$
symmetry of the model.

\section{The tree of the solutions}
\label{sec_tree}
The detailed knowledge of the temperature scenario of the states of
the mean-field SG phase is lacking so far. The main problem is that
while the more complete description available today is within the
replica approach, this task is very complex from the analytic point of
view. The TAP formalism bypasses this fundamental difficulty allowing
us for a complete description of the single sample Boltzmann states,
at least at a numerical level.

\begin{figure}
\epsfxsize=0.85\columnwidth
\epsffile{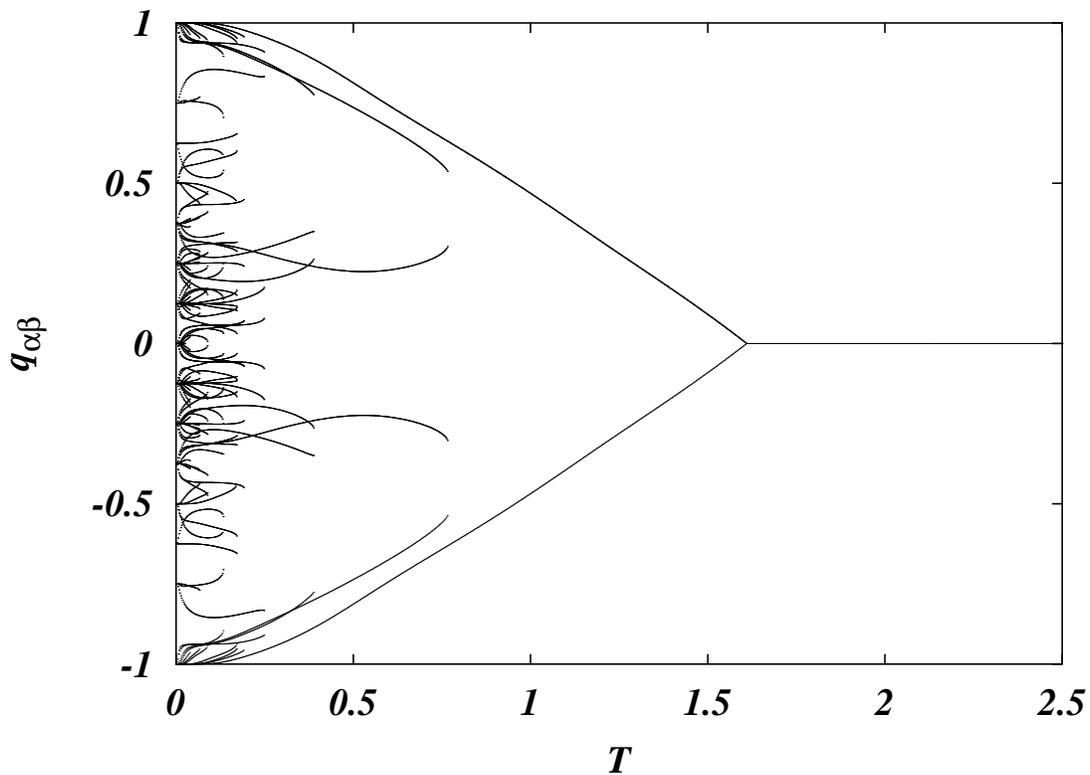}
\caption{Plot of the single sample mutual overlap of all the solutions
of \formu{eq_ntap}. Here $N=16$ and $0\leq T \leq 2.5$. The sample
critical temperature is $T_C\simeq 1.6$.}
\label{fig_trees}
\end{figure}

We have already mentioned the exponential number of solution of both
TAP equations \cite{TAP}, and the naive SK mean-field model
\cite{NEMO_TAKA_90}. It would be interesting to understand how each of
the solution is related to all the others. The similarity between
states is given by the scalar product between states, so that if
$\{m_i^\alpha(T)\}$ and $\{m_i^\beta(T)\}$ are two different states of
magnetization at a given temperature $T$ , we can define their mutual
overlap as $q_{\alpha\beta}(T)\equiv 1/N\sum m_i^\alpha(T)
m_i^\beta(T)$. Such a scalar product is directly related to the
Euclidean distance $d_{\alpha,\beta}$ of the two vectors, in fact:
\begin{equation}
d^2_{\alpha,\beta}\equiv \frac 1N \sum_{i=1}^N \left(m_i^\alpha(T)
-m_i^\beta(T)\right)^2 = q_{EA}^{\alpha}(T) + q_{EA}^{\beta}(T) - 2
q_{\alpha\beta} (T)\,.
\end{equation} 
In \fig{fig_trees} we present $q_{\alpha\beta}(T)$ calculated for
$\alpha$ and $\beta$ running through all the solutions of
\formu{eq_ntap} for a given sample of $N=16$ and all temperatures in
the window $0.0\leq T<2.5$. Decreasing the temperature from $T=2.5$ we
have a single paramagnetic solution, until we arrive around $T\sim
1.7$ which is the critical temperature of the displayed sample. In the
low temperature region until $T\sim 0.8$ we have two solutions related
by the $Z_2$ symmetry, so that their mutual overlap is just $\pm
q_{EA}(T)$. At lower temperature we find an interesting phenomenon:
the sudden appearance of another solution which is not coming from a
bifurcation of the previous ones. Let us recall that from this point
down to $T\sim 0.4$ we have $4=2\times2$ solutions, corresponding to
$6=2\times(2+1)$ points on the figure, so that in general if at a
given temperature we have $n$ solutions, $2n(n+1)/2=n(n+1)$
points will be displayed. Below $T\sim 0.4$ we see a great number of
solutions, mostly concentrated around $q=0$. The {\em merging
solutions} phenomena around $T=0$ is due to the fact that at this
temperature the local magnetizations $m_i$ have modulus $1$, therefore
the allowed values of $q_{\alpha\beta}$ are $\pm k/N$ for
$k=0,1,...,N$. The reflection symmetry of the figure with respect to
the horizontal axis is a signature of the $Z_2$ symmetry of
\formu{eq_ntap}. We have verified that all the solutions are minima of
the free energy defined in \formu{eq_free_energy}, by a numerical
check of the strict positivity of the Hessian of the matrix
$\partial_{m_i m_j}^2 F$ calculated on each solution encountered at a
given temperature. This control has been done using a standard C
\cite{NRC} on $10$ samples of $N=28$, and $100$ samples of $N=24$.

\begin{figure}
\epsfxsize=0.85\columnwidth
\epsffile{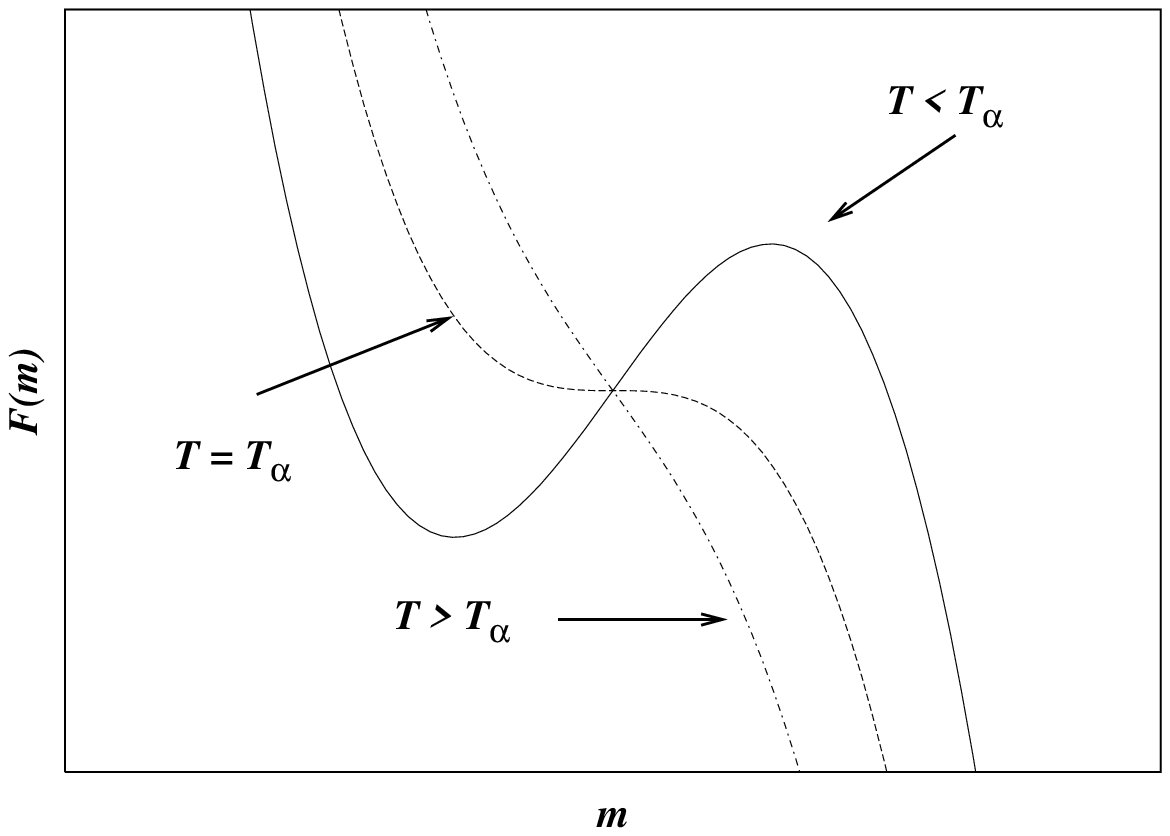}
\caption{Plot of the free energy landscape relative to the appearance
of a new solution. Here we only reproduce a schematic plot of a
one-dimensional slice of this scenario since the free energy depends on
$N$ local magnetizations.}
\label{fig_scenario}
\end{figure}

Let us now turn back our attention to the phenomenon of appearance of
solutions. The first thing we would like to understand is how this
could be interpreted in terms of the free energy landscape. Let us
denote for definiteness with $T_\alpha$, the temperature at which the
generic $\alpha$ solution appears. In principle two different
scenarios could happen: a couple of solutions merging at a certain
temperature, or the most common case according to \fig{fig_trees}, the
appearance of a single solution. While the former case can be
interpreted in terms of the usual Landau-Ginzburg potential, the
latter case can be locally described in terms of a cubic free energy
functional of the following form:
\begin{equation}
\label{eq_cubico}
F(m) \propto  m^3 + (T-T_\alpha)m^2
\end{equation}  
A consequence of this cubic potential is that around the minimum $q
\propto \sqrt{ (T-T_\alpha)}$, so that, as it is visible in
\fig{fig_trees}, the solutions appear with a sudden increase of their
derivative with respect to the temperature (See also
\fig{fig_scenario}). We also tracked some of the solutions, and we
verified that they are effectively proportional to the square root of
$T-T_\alpha$. It is interesting to note that once a solution appears
at $T_\alpha$, it can be continuously tracked down to $T=0$, which
means, from the point of view of the free energy landscape, that once
a local minimum appears, it persists down to zero temperature.

\section{The organization of the states at different temperatures}
\label{sec_org}
Given the richness of the low temperature phase space unveiled by the
tree mentioned above, it would be interesting to understand
better how these features are connected with the organization of the
states. As a first qualitative impression from \fig{fig_trees}, we
could argue that the relative distance of the free energy minima at
neighbor temperatures seem to be highly correlated, since we observe
well defined lines. From this observation we cannot conclude that the
free energy landscape changes smoothly with temperature since, at
least at the level of the tree, we are measuring the relative distance
of minima but at the same temperature, so that in principle we could
still expect a smooth covariation of minima, even in presence of
dramatic changes in temperature of the free energy landscape: the only
way to rule out this possibility is to measure the relative distance of
minima at different temperatures. 

In order to give a measure of how much temperature affects the phase
space structure, we have the following strategy:
\begin{itemize}
\item We calculate the ground state for a given sample $\{m_i(T=0)\}$.
\item We plug the ground state into the recursion equation
      \formu{eq_itera} at $T_1=\delta T = 0.01$ initialized with 
      $m_i^{(0)}= m_i(T=0)$.
\item We repeat this procedure until we arrive in the high temperature
      phase of the sample around $T=2$.
\end{itemize}
We will call this procedure {\em heating} of the ground states. In
general we can use this procedure with any state, and we can decide to
decrease instead of increasing the temperature. Hereafter we will
address to the thermal path followed by any state according to the
technique above, as {\em cooling} or {\em heating} of a state.

\begin{figure}
\epsfxsize=0.85\columnwidth
\epsffile{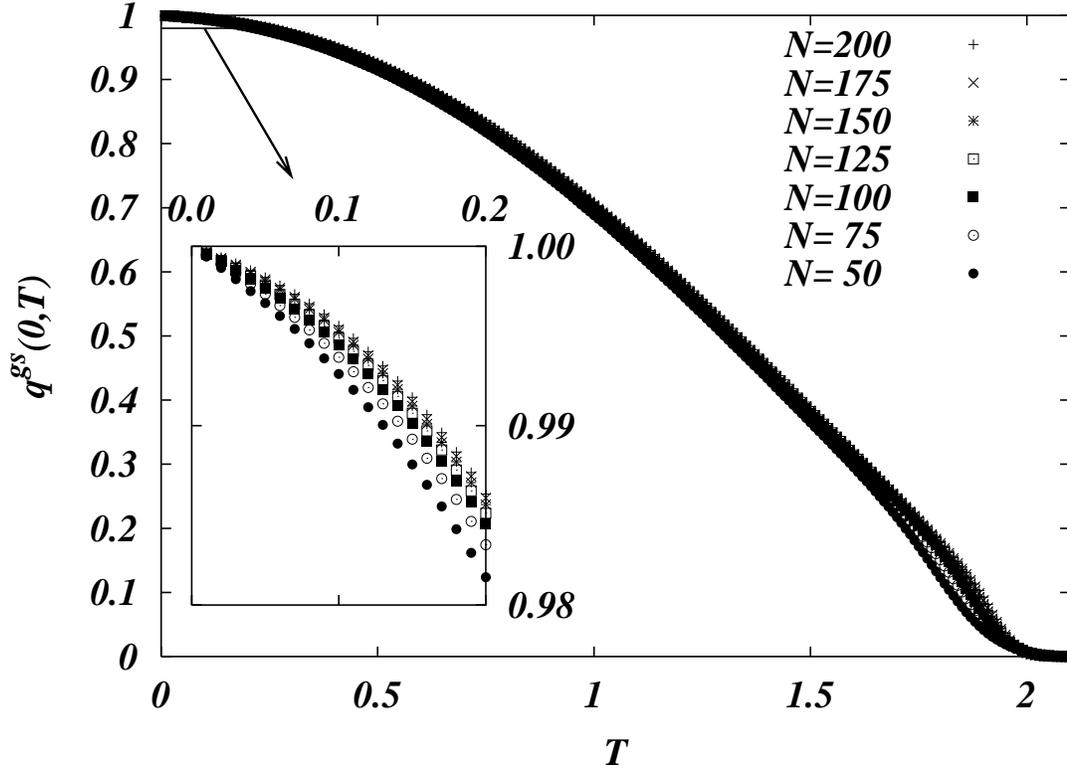}
\caption{Plot of the correlations $q^{gs}(0,T)$ defined in
\formu{eq_over0T} (see text). The details of the simulations are
displayed in \tavola{tab_salti}. Note the finite-size scaling in the
zoom in which we show a temperature window $0<T<0.2$: the bigger the
volume, the higher the correlations; the same results hold for the
whole displayed temperature range.}
\label{fig_q0T}
\end{figure}

We calculated the ground states up to volumes $N=200$ using a {\em
Reluctant} algorithm \cite{RELUCTANT}. Then we calculated the
following quantity:
\begin{equation}
\label{eq_over0T}
q_{J}^{gs}(0,T) = \frac 1N \sum_{i=1}^{N} m_i(0)m_i(T) 
\end{equation}
where we used the $\{m_i(T)\}$ obtained from the ground state heating
procedure discussed above, and we have stressed with subscript $J$ the
sample dependence of the quantity, since we will be eventually
interested in its average over the sample realizations, which we will
denote with the notation $q^{gs}(0,T)$. In \fig{fig_q0T} we display
the behavior of $q^{gs}(0,T)$, averaged over $1000$ samples and for
volume $N$ ranging from $50$ to $200$.  It is clear that the heated
ground state remains similar to its origin at $T=0$ up to the critical
temperature at $T=2$, where one has to keep in mind that $m_i(T\sim
T_C)\propto (T-T_C)^{1/2}$, so that $q^{gs}(0,T)$ above the critical
temperature is zero by definition. It is interesting to note that we
have fitted the behavior of $q^{gs}(0,T)$ near $T_C$ obtaining a well
behaved square root regime in temperature. In terms of the free energy
landscape we can conclude that the minimum originating the ground
state shifts continuously in temperature with no sign of temperature
chaos. It is interesting to point out that, as we can see from the
inset of \fig{fig_q0T}, the finite size effects even enforce this
interpretation since the higher the volume, the smaller is the effect
of the temperature shift of the minima (the same effect is present
throughout the whole low-temperature phase, as it is barely visible
from the main body of \fig{fig_q0T}). This last observation allow us
reasonably to rule out the possibility that we are observing finite
size effects.

\begin{figure}
\epsfxsize=0.85\columnwidth
\epsffile{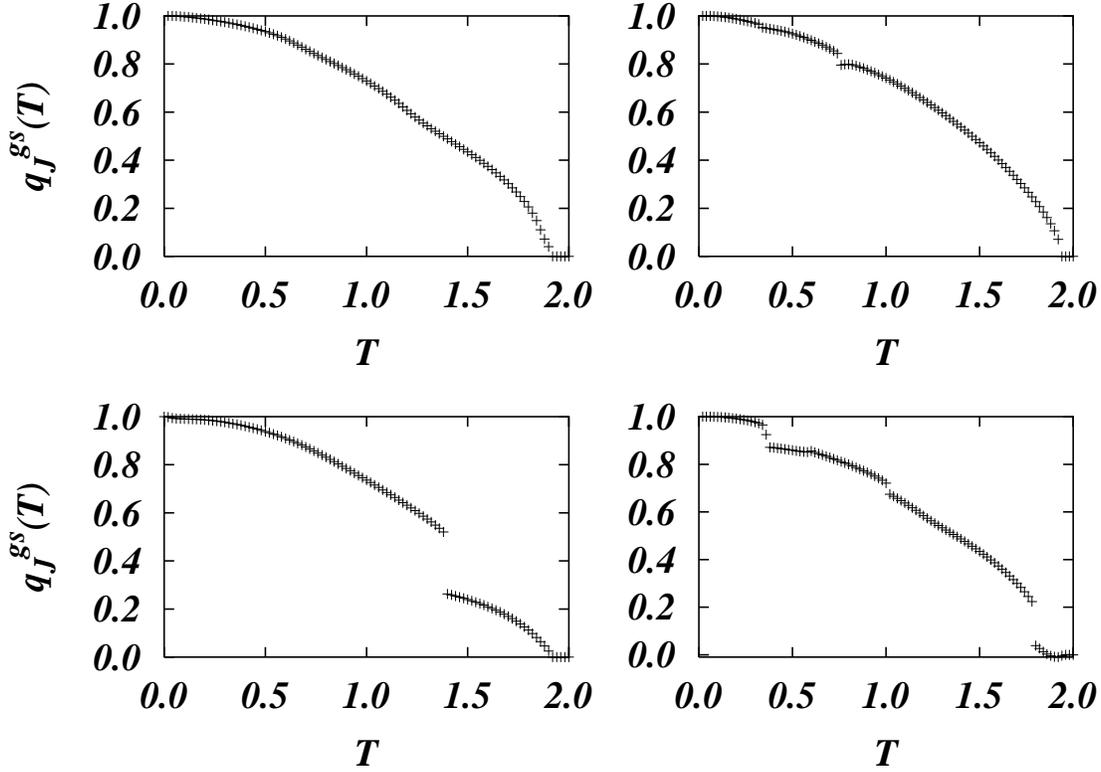}
\caption{Plot of the single sample $q^{gs}_J(0,T)$ defined in
\formu{eq_over0T}. Four different typical samples are displayed (see
text).}
\label{fig_tante}
\end{figure}

One may wonder if the disorder average actually hides something
interesting. Some of the single sample $q_{J}^{gs}(0,T)$ are displayed
in \fig{fig_tante}.  We start from the top left sample which shows the
behavior which we would expect from the sample averaged $q^{gs}(0,T)$
(see \fig{fig_q0T}), but the other three samples deserve some
comments. As the temperature is raised from $T=0$ one encounters
sudden jumps in the overlap with the ground state which can be more or
less small, but in each case the next valley selected by the
recursion, is always positively correlated with the ground state. It
is interesting to study the finite size dependence of these jumps, so
that we have defined a jump a discontinuity in the $q_J^{gs}(0,T)$
profile such that
$\left(q^{gs}_J(0,T)-q^{gs}_J(0,T+\delta)\right)>0.05$ (here
$\delta=0.01$). In \tavola{tab_salti} we show the average size of the
jumps $\Delta q^{gs}$: this quantity is almost stable with the size or
at least very slowly decreasing (at least within the volume
considered). At the same time we have also calculated the average
number of jumps per sample which turns out to be linearly increasing
with the size of the system. It is tempting to speculate a
thermodynamic limit scenario in which each sample has an infinite
number of jumps of zero size.

\begin{figure}
\epsfxsize=0.85\columnwidth
\epsffile{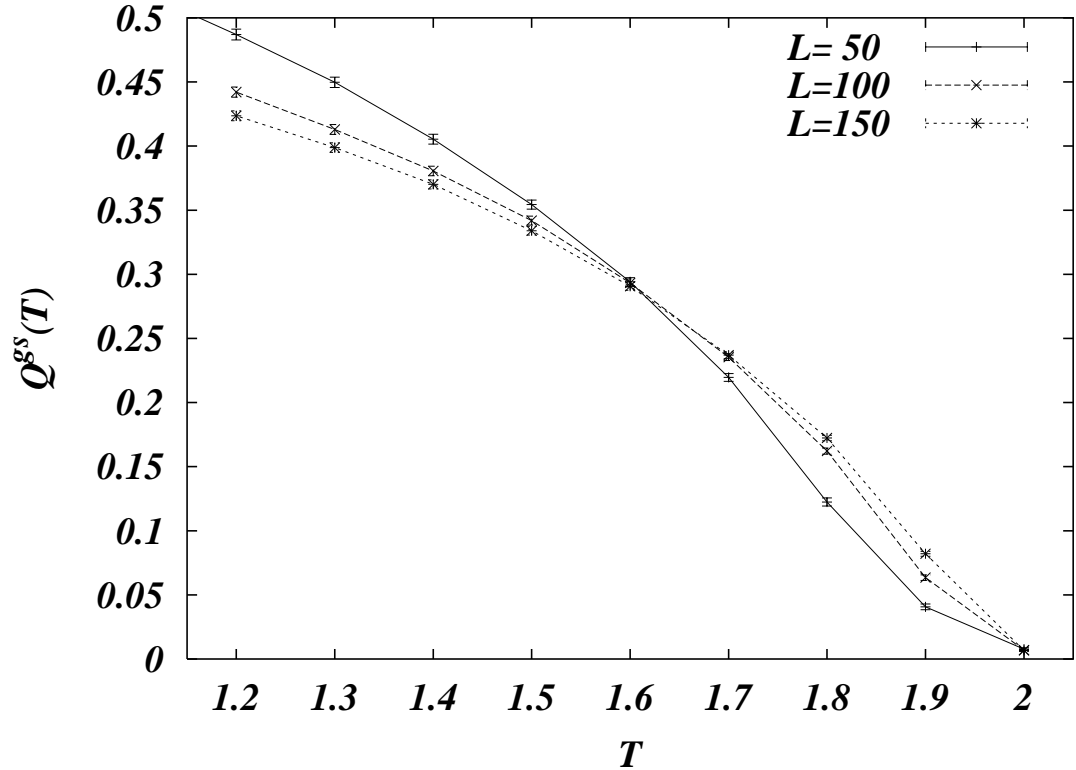}
\caption{Weighted overlap $Q^{gs}(T)$ \formu{eq_op} of the equilibrium
states at a given temperature with the ground states (see text).}
\label{fig_op}
\end{figure}

\begin{figure}
\epsfxsize=0.85\columnwidth
\epsffile{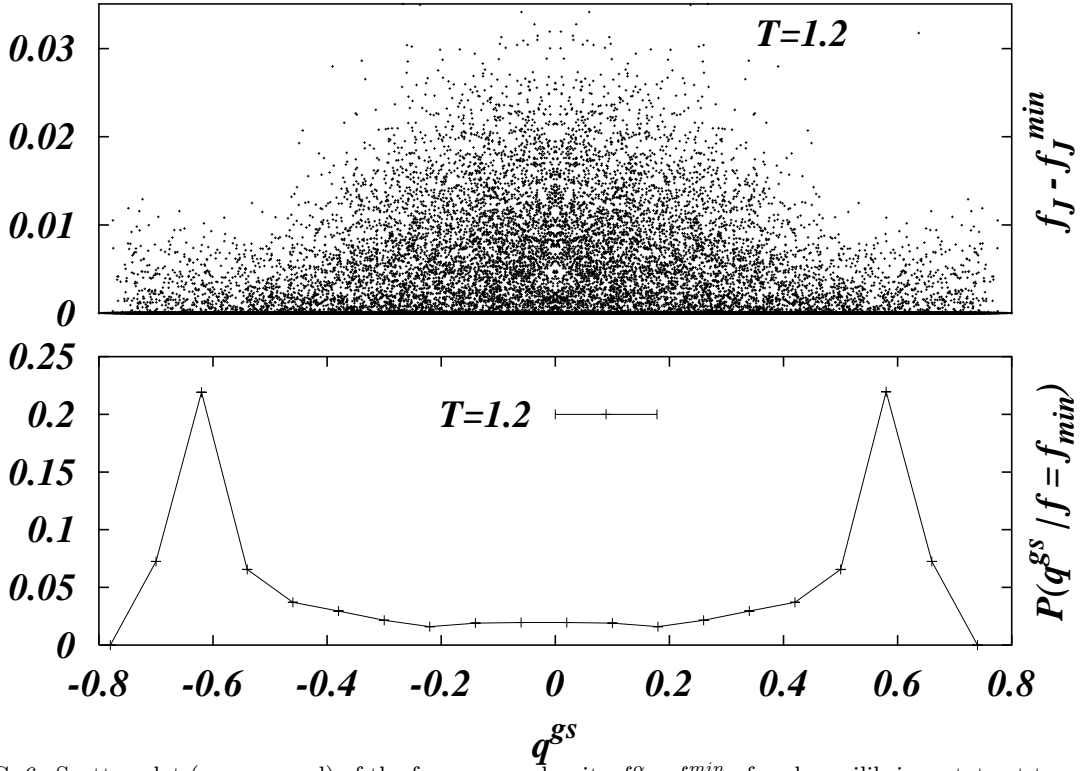}
\caption{Scatter plot (upper panel) of the free energy density
$f^{\alpha}_J-f^{min}_J$ of each equilibrium state at temperature
$T=1.2$ vs. overlap with the sample ground state
$q^{gs}_{J\alpha}(0,T)$. Here $N=150$ and the results of $1000$
different samples are superimposed. In the lower panel we display the
averaged probability distribution function defined in
\formu{pq_f_min} (see text).
}
\label{fig_scatter}
\end{figure}

So far we have concentrated our attention on the temperature
properties of the ground state free energy valley, in other words we
explored only the correlations in temperature of a single valley. One
of the advantage of working within the TAP formalism is that one can
easily get reliable information on single sample Boltzmann-Gibbs
averages. In order to give a quantitative estimate of the degree of
correlation of all the TAP states $\{m_i^\alpha(T)\}$, $\alpha =
1,...,N_S(T)$, at a certain temperature $T<T_C$ with the ground state,
we will define the following observable:
\begin{equation}
\label{eq_op}
Q^{gs}_{J}(T)  
= \sum_{\alpha=1}^{N_S(T)}w_\alpha q^{gs}_{J\alpha} (0,T) = 
\frac {\sum_{\alpha=1}^{N_S(T)} q^{gs}_{J\alpha} (0,T) e^{-N\beta f_\alpha
}}{\sum_{\alpha=1}^{N_S(T)} e^{-N\beta f_\alpha}}
\end{equation}
where $q^{gs}_{J\alpha}(0,T)=(1/N)\sum m_i(0)m^\alpha_i(T)$. Since, as
we already pointed out, for each solution $\{m^\alpha_i(T)\}$ we have
the $\{-m^\alpha_i(T)\}$ counterpart with the same free energy, in
\formu{eq_op} we considered only solutions with positive overlap with
the ground state.  We show the behavior of the disorder average of
this quantity (which we will denote as $Q^{gs}(T)$) as a function of
the temperature for sizes $N=50,100,150$ in \fig{fig_op}. Here the
considered number of samples is $1000$ for each size. Given the
relevant computational effort required in the exhaustive enumeration
of the solutions, we have limited our search in a temperature region
$T_C/2<T<T_C$. The curves show again a high degree of correlation, and
it is interesting to note that, at least below $T<1.6$, the bigger the
volumes, the lower the correlations with the ground state. We have
interpreted this phenomenon as a consequence of the appearing of a
huge number of zero-overlap solutions. It is evident that in order to
maintain such a correlated scenario, the Boltzmann-Gibbs weights of
those uncorrelated solutions must be definitely lower compared with
the correlated ones: we will try to be more precise in the following
(see discussion of \fig{fig_scatter}). Turning our attention to the
behavior of $Q^{gs}(T)$ near the critical temperature a different
finite size scenario holds: the bigger the volumes, the higher the
correlations (in this region we have for most of the samples just one
solution).

We now present some additional information about the properties of the
solutions of \formu{eq_ntap}.  In figure \fig{fig_scatter} (upper
panel) we display a scatter plot of $q^{gs}_{J\alpha} (0,T)$ vs. the
rescaled free energy density $f_J-f_J^{min}$ at $T=1.2$ calculated for
$N=1000$. The results for $1000$ samples are superimposed. The
bell-shaped pattern shows that there is a systematic correlation
between lower free energy and higher correlation with the ground
state. We think that the cloud of these uncorrelated states is a
signature of solutions appearing in deep low temperature phase. In the
lower panel of \fig{fig_scatter} we display the probability
distribution function of the lowest energy states overlaps with the
ground states averaged over $1000$ realizations of the disorder
defined by:
\begin{equation}
\label{pq_f_min}
P\left(q^{gs}|f=f_{min}\right) = 
\left[\sum_{\alpha=1}^{N_S(T)} \delta \left( q -
q^{gs}_{J\alpha}(0,T) \right)|_{f^{\alpha} = f_{min}}\right]_J\,
\end{equation}
where the square brackets denote averages over the quenched
disorder. We can observe that the distribution is strongly peaked
around $q^{gs}=\mp0.6$ confirming at a more quantitative level that
the lower the free-energy, the higher the number of states highly
correlated with the ground state is. Let us point out that a finite
size scaling analysis on the $P\left(q^{gs}|f=f_{min}\right)$ (not
shown in the plot), shows a behavior insensitive to the system size,
so that it is impossible at this level to guess the thermodynamic
limit of \formu{pq_f_min}.
\section{A heuristic algorithm for the search of the ground states}
\label{sec_algo}
The interpretation of the organization of the states at different
temperatures we have given also seems to suggest that the first
valleys appearing just below the critical temperature are also the
deepest ones down to $T=0$. This feature suggested us to try a simple
minded scheme for the search of the ground states. We start at very
high temperatures, i.e. at temperatures where we are sure that the
recursion find only the paramagnetic solutions $\{m_i=0\}$, then we
lower the temperature smoothly until we find the first state different
from zero. At this point we reproduce the heating technique introduced
above, but now decreasing the temperature until we arrive at zero. We
have compared the states obtained with this {\em cooling} scheme with
the output of the Reluctant algorithm we have used in the previous
analyses and we have verified that we obtain the same results in
average for $20\%$ of the samples.

\begin{figure}
\epsfxsize=0.85\columnwidth
\epsffile{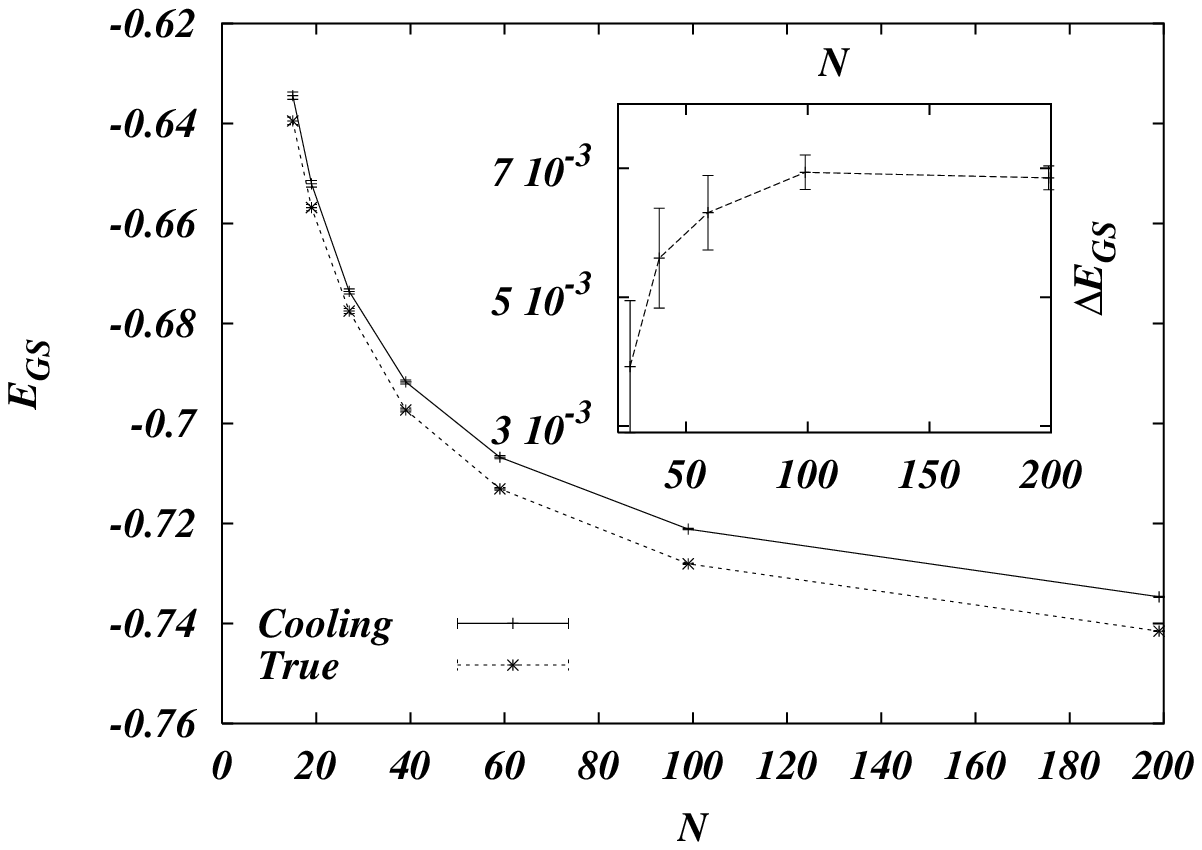}
\caption{Average ground state energy for sizes ranging from $N=19$ to
$N=199$. We label with {\em Cooling} the output of our algorithm and
with {\em True} the exact ground state. We have averaged each
point over $1000$ samples. In inset we show the difference of the two
results.}
\label{fig_py}
\end{figure}

We have implemented a systematic analysis of the performance of our
algorithm for sizes ranging from $N=19$ to $N=200$, for $10000$
samples. The results are displayed in \fig{fig_py}. They are
systematically higher out of the error bars, but it is interesting to
note that the free energy difference with the true ground states, as
showed in the inset of \fig{fig_py}, do not increases too much with
respect to the system size, lying always below $1\%$. Let us stress
that the time required by our algorithm in order to find these {\em
quasi ground-states} has a polynomial increase with time
(approximately as the square of the system size).

\begin{figure}
\epsfxsize=0.85\columnwidth
\epsffile{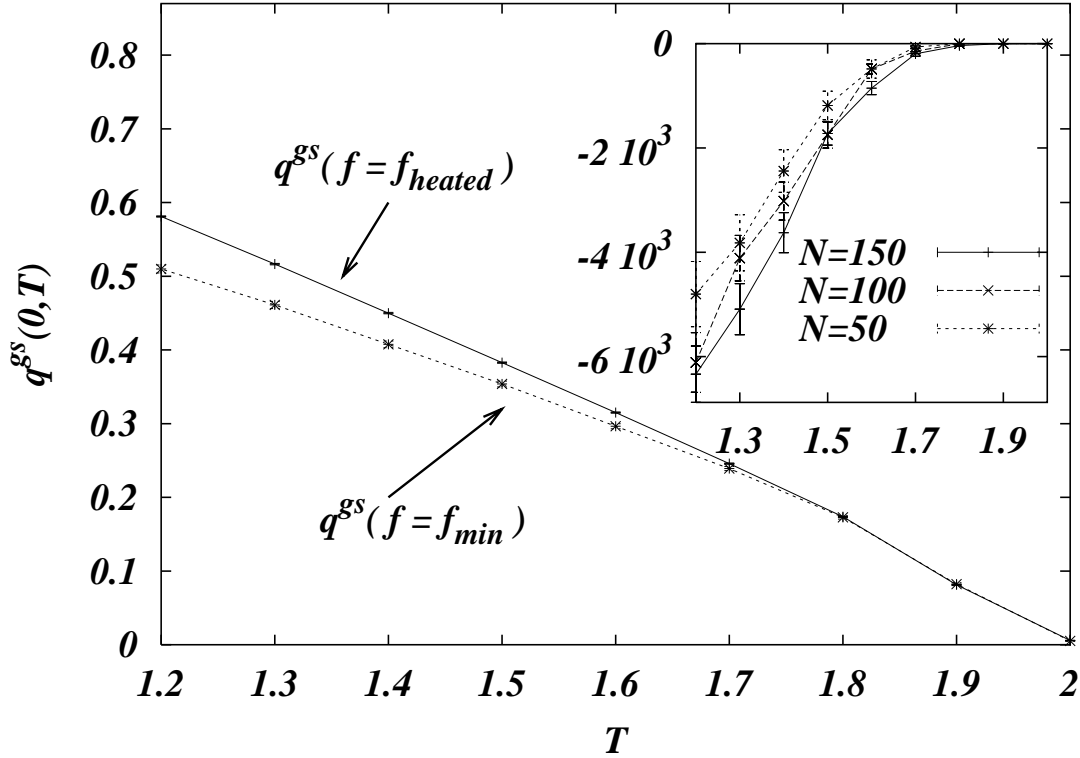}
\caption{In the main body of this figure we display the averaged
overlap $q^{gs}(0,T)$ vs. $T$. The upper curve is the same in
\fig{fig_q0T} for $N=150$ (here we have reduced the number of
considered temperatures), while the lower curve is obtained
considering only the lowest free energy states. In inset we display
the free energy difference $f_{min}-f_{heated}$ as a function of T.}
\label{fig_qfmin}
\end{figure}

The partial failure of the algorithm has suggested us a more detailed
analysis of the correlations of the true free energy minima with the
ground states. When we refer to true free energy minima, we refer to
the lowest free energy states calculated with the exhaustive
enumeration of solution previously explained.  The results are
reported in the main body of \fig{fig_qfmin} where it is shown that
these minima are always highly correlated with the ground states
(lower curve) but less correlated compared to the states obtained from
the ground-states heating procedure explained above (the upper curve
is taken from the $N=150$ curve of \fig{fig_q0T}). The results in
\fig{fig_qfmin} are obtained for $N=150$, but it seems that they do
not depend sensibly on the system size. In inset of \fig{fig_qfmin} we
show the free energy difference between states obtained by heating the
ground state, and true minima. At temperature higher than $T\sim 1.6$
we see there is no difference between this two classes of states,
while at lower temperature the heating procedure gives a free energy
difference lesser than $10^{-2}$. Note that this difference will
eventually become zero at $T=0$, and that the free energy difference
is of the same order of magnitude of the results displayed in
\fig{fig_py}.

\section{Conclusions and perspectives}
\label{sec_concl}
In this work we have studied the temperature organization of the
equilibrium states of the naive TAP mean-field equation. Our main
focus has been the problem of the chaos in temperature, i.e. whether
or not correlations between equilibrium states at different
temperatures exist. We have numerically solved the naive TAP equations
with a recursion algorithm which is able to detect only solutions
relative to the minima of the free energy functional. The equilibrium
solutions, according to our analysis, can be classified into two
families: solutions rising just below the critical temperature
following a bifurcation scenario, and solutions appearing at
temperatures well below the critical one.  We have introduced the
notion of state heating, which we have used in order to characterize
the change in temperature of the free energy landscape. We have
measured the correlation at different temperatures of the free energy
minimum relative to $T=0$ the ground state. These states are always
positively correlated throughout all the low temperature phase. The
finite size analysis of these correlations are such that we can
exclude that we are observing finite size effects. We have
measured also the weighted (according to their Boltzmann-Gibbs
thermodynamic weights) correlations of all states appearing below the
critical temperature $T_C$ with the ground state and again the
scenario suggested is highly non-chaotic. Another instructive
information gained from our numerical study is that the first minimum
appearing at $T_C$ is almost every time the deepest one. We have
exploited this unexpected feature to implement a heuristic algorithm
for the search of the ground state, just by tracking the first
solution we find at $T=T_C$ down to $T=0$. The obtained results are
interesting, since the {\em pseudo ground states} calculated with this
algorithm always are less then $1\%$ away the true minimum energy
density.

We believe that our findings support clearly a non-chaotic scenario
(at least for $N\leq 200$). The natural question is how much of this
scenario remains once we take into account also the Onsager reaction
term, i.e. in the SK model case. Let us recall that such a term goes
to zero as $T\fd 0$ so that, at sufficient low temperature, we should
find the same phenomenology. Moreover it is known that the naive TAP
equation shows full replica symmetry breaking (RSB), much the same the
SK model, another hint in the sense that what we observe it might be
common to all mean-field model with RSB scenario. However it would be
interesting to define the same notion of state temperature evolution,
for the full fledged TAP equations \formu{eq_tap}. But if on one hand
we are confident that, at least at the mean field level, the
non-chaotic scenario seems to be likely enough, on the other hand we
believe that chaos in short range systems remains the most challenging
problem, and an extension of our techniques for the Edwards-Anderson
problem should be instructive.

We acknowledge interesting discussions with T.~Rizzo, A.~Cavagna,
A.~Crisanti, and I.~Giardina. We especially thank E.~Marinari and
M.~Ratieville for interesting discussions and a careful reading of the
manuscript. We would like also to thank M.~Palassini and A.P.~Young
for providing us the first set of exact ground states. R.M. thanks the
interchange program between the University of Rome ``La Sapienza'' and
the University of ``La Habana''. A.P. acknowledges the kind
hospitality of ICTP where part of this work was done. Part of the
simulations were carried out on the Pentium cluster of the University
of Cagliari {(\em Kalix 2)} funded by Italian MURST 1998 COFIN.

\vfill
\begin{table}
\begin{tabular}{clcc}
$N$ & $\Delta q^{gs}$ & $\#\mathrm{jumps}$ & $\# \mathrm{samples}$ \\\hline
50  & 0.149(4)  & 0.78(1) &  10000 \\
75  & 0.154(4)  & 0.82(1) &  10000 \\
100 & 0.150(4)  & 0.86(1) &  10000 \\
125 & 0.149(4)  & 0.90(1) &  10000 \\
150 & 0.147(4)  & 0.95(1) &  10000 \\
175 & 0.144(8)  & 0.97(1) &  20000 \\
200 & 0.14(1)   & 1.00(2) &  17186 
\end{tabular}
\caption{In this table we display the results for the average size
$\Delta q^{gs}$ of the jumps (second column), and the average number
of jumps per sample (third column). The number of samples considered
is in the fourth column.}
\protect\label{tab_salti}
\end{table}


\begin{thebibliography}{99}

\bibitem[1]{rm} Email: {\tt mulet@ictp.trieste.it} Permanent Address:
Superconductivity Laboratory, Physics Faculty - IMRE, University of
Havana, CP 10400, La Habana, Cuba.
\bibitem[2]{ap} Email: {\tt andrea.pagnani@roma1.infn.it}
\bibitem[3]{gp} Email: {\tt giorgio.parisi@roma1.infn.it}

\bibitem{BOOKS} 
   K. Binder and A. P. Young,
   Rev. Mod. Phys. {\bf 58}, 801 (1986);
   M. M\'ezard, G. Parisi and M. A. Virasoro, 
   {\em Spin Glass Theory and Beyond} 
   (World Scientific, Singapore 1987);
   K. H. Fisher and J. A. Hertz,
   {\em Spin Glasses}
   (Cambridge University Press, Cambridge U.K. 1991).

\bibitem{TAP}
   D.J. Thouless, P.W. Anderson, and R.G. Palmer, 
   Phil. Mag. {\bf 35}, 593 (1987).

\bibitem{BRAY_MOORE}
   A.J. Bray, and M.A. Moore,
   J. Phys. C: Solid State Phys. {\bf 12} L441 (1979);	
   {\em ibid.} {\bf 13} L469 (1980). 	

\bibitem{NEMO_TAKA_85}
   K. Nemoto, and H. Takayama, 
   J. Phys. C: Solid State Phys. {\bf 18} L529 (1985).

\bibitem{BRAY_SOMPO_YU}
   A.J. Bray, H. Sompolinsky, and C. Yu, 
   J. Phys. C: Solid State Phys. {\bf 19} 6389 (1986).

\bibitem{NEMO_TAKA_90}
   H. Takayama, and K. Nemoto,
   J. Phys.: Condens. Matter {\bf 2} 1997 (1990). 

\bibitem{LA_MA_PA} 
   M. Guagnelli, E. Marinari, and G. Parisi,
   J. Phys. A: Math. Gen. {\bf 26} 5675 (1993); 	
   D. Lancaster, E. Marinari, and G. Parisi,	
   {\em ibid.} {\bf 28}, 3959 (1995).

\bibitem{FISHER_HUSE_86}
   D.S. Fisher and D.A. Huse,
   Phys. Rev. Lett. {\bf 56} 1601 (1986).

\bibitem{BRAY_MOORE_87}
   A.J. Bray and M.A. Moore,
   Phys. Rev. Lett. {\bf 58}, 57 (1987).

\bibitem{FISHER_HUSE_88}
   D.S. Fisher and D.A. Huse,
   Phys. Rev. B {\bf 38}, 373 (1988).

\bibitem{EXPERIMENTS}
   K. Jonason, E. Vincent, J. Hammam, J.P. Bouchaud, and P. Nordblad,
   Phys. Rev. Lett. {\bf 81}, 3243 (1998).

\bibitem{REAL_SPACE1} 
   J.-P. Bouchaud 
   in {\em Soft and Fragile Matter}, ed: M.E. Cates and M.R. Evans 
   (Institute of Physic Publishing, Bristol and Philadelphia, 2000). 

\bibitem{REAL_SPACE2}
   L.F. Cugliandolo and J. Kurchan,
   Phys. Rev. B {\bf 60}, 922 (1999).

\bibitem{RIEGER}
   H. Rieger, L. Santen, U. Blasum, M. Diehl, and M. J\"unger,
   J. Phys. A: Math. Gen. {\bf 29} 3939 (1996). 

\bibitem{KONDOR_1}
   I. Kondor,
   J. Phys. A: Math. Gen. {\bf 22}, L163 (1989).

\bibitem{KONDOR_2}
   I. Kondor and A. V\'egs\"o,
   J. Phys. A: Math. Gen. {\bf 26}, L641 (1993). 

\bibitem{FRANZ_NEY-NIFLE} 
   S. Franz and M. Ney-Nifle,
   J. Phys. A: Math. Gen. {\bf 28}, 2499 (1995).

\bibitem{FRA_PA_VI}
   S. Franz, G. Parisi and M.A. Virasoro,
   J. Physique I {\bf 2}, 1969 (1992).

\bibitem{TESI_TOMMASO} 
   T. Rizzo, 
   {\em Modelli di vetri di spin in campo medio: caos in temperatura}, 
   Tesi di Laurea, Universit\`a di Roma ``La Sapienza'' (1999).


\bibitem{FELIX} 
   F. Ritort,
   Phys. Rev. B {\bf 50}, 6844 (1994).

\bibitem{NEY}
   M. Ney-Nifle, 
   Phys. Rev. B {\bf 57}, 492 (1998).

\bibitem{NEY_YOUNG}
   M. Ney-Nifle and A.P. Young,
   J. Phys. A: Math. Gen. {\bf 30}, 5311 (1997). 

\bibitem{KISKER}
   J. Kisker, L. Santen, M. Schreckenberg, and H. Rieger 
   Phys. Rev. B {\bf 53} 6418 (1996). 


\bibitem{BILL_MARI}
   A. Billoire and E. Marinari,
   J. Phys. A: Math. Gen. {\bf 33}, L265 (2000).   

\bibitem{NISHI_NEMO_TAKA}
   K. Nishimura, K. Nemoto, and H. Takayama,
   J. Phys. A: Math. Gen. {\bf 23} 5915 (1990).

\bibitem{MEHTA}
   M.L. Mehta, 
   {\em Random Martices and the Statistical Theory of
   Energy Levels}, (Academic Press, New York, 1991). 


\bibitem{DE_DOM_YOUNG}
   C. De Dominicis and A.P. Young,
   J. Phys. A: Math. Gen. {\bf 16} 2063 (1983).

\bibitem{FIRST_RECUR}
   H. Yoshizawa, and D.P. Belanger, 
   Phys. Rev. B {\bf 30} 5220 (1984);
   C. Ro, G.S. Grest, C.M. Soulolis, and K. Levin,
   Phys. Rev. B {\bf 31} 1682 (1985). 

\bibitem{NRC}
   W.H. Press, S.A. Teukolsky, W.T. Wetterling, and B.P. Flannery, 
   {\em Numerical Recipes in C. The Art of Scientific Computing},
   (Cambridge Univeristy Press, 1992).
 
\bibitem{RELUCTANT} 
   The Reluctant algorithm works as follows: We generate random 
   $\{m_i(0)\}$, then we calculate the local magnetic fields
   $h_i= \sum_j J_{ij}m_j$. We are interested in reducing to zero the
   number of spins opposite to the local fields, so that as a
   {\em Reluctant} strategy each time we flip the spin relative to the 
   smaller (with regard to its absolute value) local field $h_i$,
   then we calculate again the local fields until 
   all spins are aligned with their local fields. We iterate this
   procedure until we find for $10$ times the same lower energy state. 
   This procedure was introduced by G.~Parisi in {\em On the
   Statistical Properties of the Large Time Zero Temperature Dynamics
   of the SK Model}, {\tt cond-mat/9501045}.  
     
\end{thebibliography}
\end{document}